\documentstyle[11pt,epsfig]{article}

\def\Journal#1#2#3#4{{#1} {\bf #2}, #3 (#4)}

\def\NPB{{\em Nucl. Phys.} B}
\def\PLB{{\em Phys. Lett.}  B}
\def\PRL{\em Phys. Rev. Lett.}
\def\PRD{{\em Phys. Rev.} D}

\def\be{\begin{equation}}
\def\ee{\end{equation}}
\def\bea{\begin{eqnarray}}
\def\eea{\end{eqnarray}}

\begin{document}

\hspace*{\fill}\parbox[t]{2.9cm}{
BNL-HET-00/28}
\vspace{1.cm}
\begin{center}
{\Large\bf Hadronic Three Jet Production at Next-to-Leading
Order}\footnote{Talk presented by W.B.K. at the XXXVth Rencontres de
Moriond: QCD and High Energy Hadronic Interactions, Les Arcs, France,
March 18--25, 2000.} \\
\vspace{1.cm}
{ William B. Kilgore }
\\
\vspace{.2cm}
{\sl Department of Physics\\
Brookhaven National Laboratory\\
Upton, New York 11973-5000, USA}\\
\vspace{.4cm}
{ Walter T. Giele }
\\
\vspace{.2cm}
{\sl Theoretical Physics Department\\
Fermi National Accelerator Laboratory\\
Batavia, Illinois 60510, USA}\\
\begin{abstract}
I present results of a next-to-leading order calculation of three jet 
production at hadron colliders.  This calculation will have many
applications.  In addition to computing three-jet observables
(spectra, mass distributions),  this calculation permits the first
next-to-leading order studies (at hadron colliders) of jet and event
shape variables. 
\end{abstract}
\end{center}

\section{Introduction}
One of the difficulties in interpreting experimental results is
in assessing the uncertainty to be associated with the theoretical
calculation.  In QED and the weak interactions, one generally has
confidence in the accuracy of leading order (LO) calculations because
the couplings are sufficiently weak that higher order corrections are
small.  In QCD, however, the coupling is quite strong and it is
difficult to obtain a reliable estimate of the theoretical
uncertainty.

One typically characterizes theoretical uncertainty by the dependence
on the renormalization scale $\mu$. Since one doesn't actually know how
to choose $\mu$ or even a range of $\mu$, the uncertainty associated
with scale dependence is somewhat arbitrary.  One motivation for
performing next-to-leading order (NLO) calculations is to reduce the
scale dependence associated with the calculation.

However, this is not the only benefit of an NLO calculation.  There
are times when the LO calculation is a bad estimator of the
physical process.  It may be that leading order kinematics
artificially forbids the most important physical process.  It could
also be that the NLO corrections are simply large.  
Even if the overall NLO correction is relatively small, there may be
regions of phase space, where NLO corrections are large.  It is
only in those regions of phase space where the NLO corrections are
well behaved (as determined by the ratio of the NLO to LO terms) that
one has confidence in the reliability of the calculation and can begin
to believe the uncertainty estimated from scale dependence and it is
only when one has a reliable estimate of the theoretical uncertainty
that comparisons to experiment are meaningful.

\section{Methods}
The NLO three jet calculation consists of two parts: two to three
parton processes at one-loop (the virtual terms) and two to four
parton processes (the real emission terms) at tree-level.  Both of
these contributions are infrared singular; only the sum of the two is
infrared finite and meaningful.  The virtual contributions are
infrared singular because of loop momenta going on-shell.  The
real emission contributions are singular when two partons become
collinear or when a gluon becomes very soft.  The
Kinoshita-Lee-Nauenberg theorem~\cite{kln} guarantees that the
infrared singularities cancel for sufficiently inclusive processes
when the real and virtual contributions are combined.

The parton sub-processes involved are $gg\to ggg$~\cite{BDK1},
$\overline{q}q\to ggg$~\cite{BDK2},
$\overline{q}q\to\overline{Q}Qg$~\cite{KST}, and processes related to 
these by crossing symmetry, all computed to one-loop, and $gg\to
gggg$, $\overline{q}q\to gggg$, $\overline{q}q\to\overline{Q}Qgg$, and
$\overline{q}q\to\overline{Q}Q\overline{Q^\prime}Q^\prime$ and the
crossed processes computed at tree-level.

In order to implement the kinematic cuts necessary to compare a
calculation to experimental data one must compute the cross section
numerically.  Thus, it is not sufficient to know that the
singularities drop out in the end, we must find a way of canceling
them before we start the calculation.  Several different
methods of implementing this infrared
cancellation  have been successfully employed in various NLO
calculations. The method we use is the ``subtraction improved'' phase
space slicing method~\cite{KG}.  Phase space slicing~\cite{GG,GGK} 
uses a resolution criterion $s_{\rm min}$, which
is a cut on the two parton invariant masses,
\begin{equation}
s_{ij} = 2E_iE_j(1-\cos\theta_{ij}).
\end{equation}
If partons $i$ and $j$ have $s_{ij}>s_{\rm min}$ they are said to
be resolved from one another.  (Which is not to say that a jet
clustering algorithm will not put them into the same jet.)  If
$s_{ij}<s_{\rm min}$ partons $i$ and $j$ are said to be unresolvable.
One advantage of the $s_{\rm min}$ criterion is that it simultaneously
regulates both soft ($E_i\to0$ or $E_j\to0$) and collinear
($\cos\theta_{ij}\to1$) emission.  In the rearrangement of terms, the
infrared region of phase space is where any two parton invariant mass
is less than $s_{\rm min}$.  These regions are sliced out of the full
two-to-four body phase space, partially integrated and then added to
the two-to-three body integral.

Because the infrared integral is bounded by $s_{\rm min}$, both the
two-to-three and two-to-four body integrations are logarithmically
dependent on $s_{\rm min}$.  Since $s_{\rm min}$ is an arbitrary
parameter the sum of the two contributions must be $s_{\rm min}$
independent.  Thus, we have rearranged the calculation, trading a
cancellation of infrared poles for a cancellation of logarithms of
$s_{\rm min}$.  The demonstration of $s_{\rm min}$ independence
implies that we have correctly implemented the infrared cancellation. 

\section{Results}
The results shown below were computed for the following kinematic
conditions: the $\bar{p}p$ center of mass energy was $1800\ {\rm
GeV}$; at least one jet was required to have more than $100\ {\rm
GeV}$ of transverse energy ($E_T$), while two more jets were required
to have more than $50\ {\rm GeV}$ of transverse energy.  All three
jets were required to lie in the pseudorapidity range $-4.0 < \eta_J <
4.0$.

\begin{figure}
\hbox to \hsize{\hfil  \epsfxsize\hsize\epsfbox{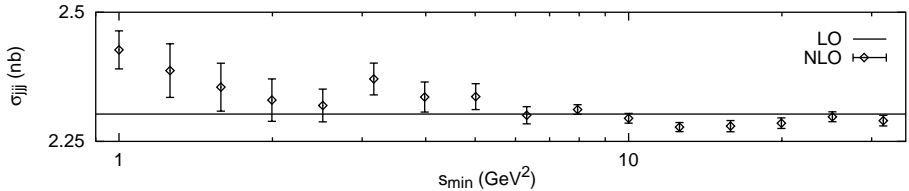}\hfil}   
\caption{\small
Next-to-leading order three jet cross section vs. $s_{\rm
min}$. The leading order cross section is shown as a solid line.
}
\label{fig:sminplot}
\end{figure}

The first test of the calculation is to demonstrate $s_{\rm min}$
independence of the cross section.  In figure~\ref{fig:sminplot}, the
next-to-leading order cross section is computed for sixteen values of
$s_{\rm min}$ between $1$ and $40\ {\rm GeV}^2$.  We see that the
computed cross section is stable over a wide range of $s_{\rm min}$.
The next-to-leading order calculations were all performed using {\sc
CTEQ3M} parton distributions.  The renormalization and factorization
scales were chosen to be $\mu=100\ {\rm GeV}$.

For comparison, the leading order calculation (computed using the {\sc
CTEQ3L} parton distributions but all other parameters the same) is
shown as a solid line.  We see that for this choice of parameters, the
magnitude of the next-to-leading correction is small.

\begin{figure}[ht]
\hbox to \hsize{\hfil \epsfxsize\hsize\epsfbox{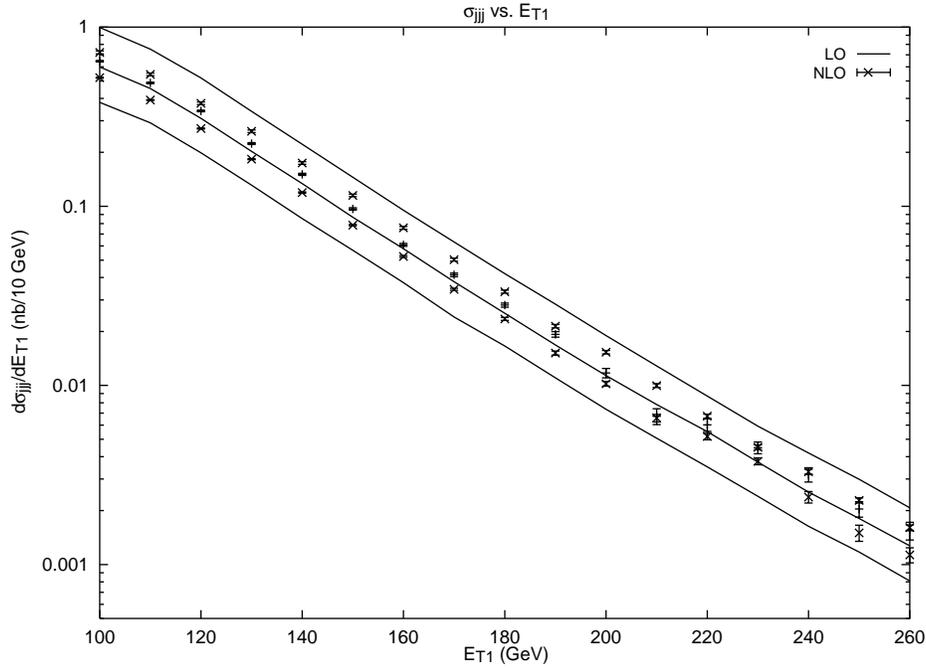}\hfil}   
\caption{
\small Transverse energy spectrum of the leading jet for different
choices of scales.  The next-to-leading order results are shown as
points and the leading order results as solid lines.  In each case,
the upper points correspond to $\mu=E_{T1}/2$, the middle points to
$\mu=E_{T1}$ and the lower points to $\mu=2E_{T1}$.}
\label{fig:ET1plot}
\end{figure}

I have also computed the transverse energy spectrum of the leading jet
($E_{T1}$).  Figure~\ref{fig:ET1plot} shows a first attempt to explore
the scale dependence of the calculation.  The kinematic cuts (and
parton distributions) for the results shown in
figure~\ref{fig:ET1plot} are the same as in figure~\ref{fig:sminplot},
but the renormalization and factorization scales for the upper, middle
and lower curves are chosen such that $\mu_F = \mu_R = E_{T1}/2,
E_{T1}, 2E_{T1}$ respectively.  For the next-to-leading order
calculation, $s_{\rm min}$ was chosen to be $\sim20\ {\rm GeV}^2$.

Comparing to the leading order results (shown as solid lines), we see
that the next-to-leading corrections are indeed small, and that the
scale dependence is substantially reduced.

\section{Applications}
The next-to-leading order calculation of three jet production will
have a wide array of phenomenological applications.
\subsection{Measurement of $\alpha_s$}
It should be possible to extract a purely hadronic measurement of
$\alpha_s$.  One possibility for such a measurement would be a
comparison of the three jet to two jet event rate.  Since both
processes are sensitive to all possible initial states at tree-level,
a next-to-leading order comparison should be relatively free of bias
from the parton distributions.  Because the measurement will be
simultaneously performed over a wide range of energy scales, the
running of $\alpha_s$ can be used to constrain the fits and enhance
the precision of the combined measurement.
\subsection{Study jet clustering algorithms}
Because there are up to four partons in the final state, as many as
three partons can end up in a single jet.  This makes the three jet
calculation sensitive to the details of jet clustering algorithms.
This sort of study in pure gluon production~\cite{KG} uncovered an
infrared sensitivity in the commonly used iterative cone algorithms.
\subsection{Study jet structure and shape}
Because there can be three partons clustered into a single jet, this
calculation will allow truly next-to-leading order studies of the
energy distribution in jets.  Studies of jet production in deep
inelastic scattering~\cite{KRRZ} show that the next-to-leading order
correction for this variable is substantial and agrees rather well
with experimental measurements.
\subsection{Study event shape variables}
There has been a long history of studying event shape variables like
Thrust at $e^+e^-$ colliders.  These measurements challenge the
ability of perturbative QCD to describe the data and provide a
means (other than event rate) of obtaining a precise measurement of
$\alpha_s$.  It will be interesting to see if one can make a
meaningful study of such variables at hadron colliders.

\section*{Acknowledgments}
This work was supported by the US Department of Energy under grant
DE-AC02-98CH10886.


\begin{thebibliography}{99}
\bibitem{kln}
T.~Kinoshita, \Journal{{\em J. Math. Phys.}}{3}{650}{1962};\\
T.D.~Lee and M.~Nauenberg, \Journal{{\em Phys. Rev.}}{133}{1549}{1964}.

\bibitem{BDK1}
Z.~Bern, L.~Dixon, D.A.~Kosower, \Journal{\PRL}{70}{2677}{1993}\
[hep-ph/9302280].

\bibitem{BDK2}
Z.~Bern, L.~Dixon and D.A.~Kosower, \Journal{\NPB}{437}{259}{1995}\
[hep-ph/9409393].

\bibitem{KST}
Z.~Kunszt, A.~Signer and Z.~Tr\'ocs\'anyi,
\Journal{\PLB}{336}{529}{1994}\ [hep-ph/9405386].

\bibitem{GG}
W.T. Giele and E.W.N Glover, \Journal{\PRD}{46}{1980}{1992}.

\bibitem{GGK}
W.T. Giele, E.W.N. Glover and D.A. Kosower,
\Journal{\NPB}{403}{633}{1993}\hfil\break [hep-ph/9302225].

\bibitem{KG}
W.B. Kilgore and W.T. Giele, \Journal{\PRD}{55}{7183}{1997}.

\bibitem{KRRZ}
N. Kauer, L Reina, J. Repond, and D. Zeppenfeld,
\Journal{PLB}{460}{189}{1999}\ [hep-ph/9904500].

\end{thebibliography}
\end{document}